\documentclass[]{revtex4}
\usepackage{amsmath}
\usepackage{amsfonts}
\usepackage{amssymb}
\usepackage{hyperref}
\usepackage{graphicx}
\usepackage{amsthm}
\usepackage{color}
\usepackage{float}
\usepackage{wrapfig}
\usepackage{epsfig}
\usepackage{natbib} 

\newtheorem*{thm*}{Theorem}

\begin{document}

\title{Algebraic $\mathcal{L}_{q}$-norms and complexity-like properties of Jacobi polynomials-Degree and parameter asymptotics.}

\author{Nahual Sobrino}
\address{Donostia International Physics Center, Paseo Manuel de Lardizabal 4, E-20018 San Sebasti\'an, Spain}
\address{Nano-Bio Spectroscopy Group and European Theoretical Spectroscopy Facility (ETSF), Departamento de Pol\'imeros y Materiales Avanzados: F\'isica, Qu\'imica y Tecnolog\'ia, Universidad del Pa\'is Vasco UPV/EHU, Avenida de Tolosa 72, E-20018 San Sebasti\'an, Spain}

\author{Jes\'us S. Dehesa}
\email{dehesa@ugr.es}
\affiliation{Instituto Carlos I de F\'{\i}sica Te\'orica y Computacional, Universidad de Granada, Granada 18071, Spain}
\affiliation{Departamento de F\'{\i}sica At\'{o}mica, Molecular y Nuclear, Universidad de Granada, Granada 18071, Spain}

\begin{abstract} 
 
 The Jacobi polynomials  $\hat{P}_n^{(\alpha,\beta)}(x)$ conform the canonical family of hypergeometric orthogonal polynomials (HOPs) with the two-parameter weight function $(1-x)^\alpha (1+x)^\beta, \alpha,\beta>-1,$ on the interval $[-1,+1]$. The spreading of its associated probability density (i.e., the Rakhmanov density) over the orthogonality support has been quantified, beyond the dispersion measures (moments around the origin, variance), by the algebraic $\mathfrak{L}_{q}$-norms (Shannon and R\'enyi entropies) and the monotonic complexity-like measures of Cram\'er-Rao, Fisher-Shannon and LMC (L\'opez-Ruiz, Mancini and Calbet) types. These quantities, however, have been often determined in an analytically highbrow, non-handy way; specially when the degree or the parameters $(\alpha,\beta)$ are large. In this work, we determine in a simple, compact form the entropic and complexity-like properties of the Jacobi polynomials in the two extremal situations:  ($n\rightarrow \infty$; fixed $\alpha,\beta$) and  ($\alpha\rightarrow \infty$; fixed $n,\beta$). These two asymptotics are relevant \textit{per se} and because they control the physical entropy and complexity measures of the high energy (Rydberg) and high dimensional (pseudoclassical) states of numerous supersymmetric quantum-mechanical systems.

\end{abstract}


\maketitle

\section{Introduction}


The formalization of the intuitive notion of simplicity/complexity of probability distributions is a formidable task, not yet well established in spite of a huge number of efforts in many scientific and technological areas from quantum chemistry  \cite{Sen2011,Antolin2009,Badii1997,Shiner1999} and quantum technologies \cite{Nielsen2000,Cubitt2014,Bruss2019,Vitanyi2001,Lloyd2001} to applied mathematics and approximation theory \cite{Dehesa2006,Dehesa2015,Ghosh2019,Rudnicki2016}. To a large extent, this is because of the great diversity of configurational shapes from perfect order to  maximal randomness (or perfect disorder) which have the bi- and multi-parametric probability distributions associated to the quantum one- and many-body systems and the special functions of applied mathematics and mathematical physics \cite{Nikiforov1988}. This enormous amount of geometrical forms cannot be captured by a single quantity, but it requires a number of measures of intrinsic and extrinsic characters. The latter ones refer to algorithmic and computational complexities \cite{Goldreich2008,Kaltchenko2021} which are  closely related to the time required for a computer to solve a given problem, so that they depend on the chosen computer. The former ones refer to statistical measures of complexity, extracted from the density functional theory of electronic systems \cite{Parr1989}, which quantify the degree of structure or pattern of one- and many-electron systems in terms of the single electron density. The main quantities of this type are the Cram\'er-Rao \cite{Dehesa2006,Antolin2009}, Fisher-Shannon \cite{Vignat2003,Angulo2008,Romera2004} and LMC (L\'opezruiz, Mancini and Calbet) \cite{Lopezruiz1995,Catalan2002} complexities and modifications of them  \cite{Pipek1997,Lopez2009,Lopez2005,Romera2008,Romera2009,Sanchez2014,Martin2003,LopezRosa2009,Toranzo2017,Puertas2017c,Puertas2017d}. Here we will use the basic density-dependent complexity measures (Cram\'er-Rao, Fisher-Shannon and LMC) recently introduced in electronic structure (see e.g.  \cite{Martin2006,Dehesa2009,LopezRosa2009,Molina2012} and the reviews \cite{Angulo2011,Dehesa2011}), which are of intrinsic character in the sense that they do not depend on the context but on the probability density of the system under consideration. These quantities are given by the product of two spreading measures of dispersion (variance) and entropic (Fisher information, Shannon entropy, R\'enyi entropy) types, so that each measures two configurational shapes of the system in a simultaneous manner.\\

The hypergeometric polynomials, $\{p_n(x)\}$, orthogonal with respect to the weight function $h(x)$ on the support  interval $\Lambda$, are known to often control the quantum-mechanical wavefunctions of the bound states in numerous quantum systems \cite{Nikiforov1988,Bagrov1990,Cooper2001,Chaitanya2012,Yanez1994,Gangopadhyaya2017}. The Hermite, Laguerre and Jacobi polynomials are the three canonical families of real hypergeometric orthogonal polynomials (HOPs) \cite{Andrews1999,Ismail2005,Olver2010,Koekoek2010}. Recently the entropy-  and complexity-like properties of these polynomials, which determine their spreading on the orthogonality interval, have begun to be investigated by means of the entropy- and complexity-like measures \cite{Dehesa2015,Dehesa2021,Sobrino2021} of the associated Rakhmanov density $\rho_n(x) =  p_n^2(x)\,h(x)$. This normalized-to-unity probability density function governs the ($n\to+\infty$)-asymptotics of the ratio of two polynomials with consecutive orders \cite{Rakhmanov1977}, and characterizes the Born's probability density of the bound stationary states of a great deal of quantum-mechanical potentials which model numerous atomic and molecular systems \cite{Nikiforov1988,Gangopadhyaya2017,Ghosh2019,Alhaidari2021,Assi2021,Nath2021}. The numerical evaluation of the integral functionals corresponding to the dispersion, entropic and complexity measures of the HOPs by means of the standard quadratures is not convenient, because the highly oscillatory nature of the integrand renders Gaussian quadrature ineffective as the number of quadrature points grows linearly with $n$ and the evaluation of high-degree polynomials are subject to round-off errors. Indeed, since all the zeros of $p_n$ belong to the interval of orthogonality, the increasing number of integrable singularities spoil any attempt to achieve reasonable accuracy even for rather small $n$ \cite{Buyarov2004,Perrey2009}.\\

The entropic and complexity-like measures of the three canonical families of the HOPs have been analytically calculated in terms of the degree and the parameters which characterize their weight function. The resulting analytical expressions for the Cram\'er-Rao complexity of Hermite and Laguerre polynomials are simple and compact \cite{Dehesa2015}, but for the rest of complexity measures of HOPs this is not at all true because the involved entropic components have a somewhat highbrow, non-handy form, being mostly useful in an algorithmic sense \textit{only} (see the review \cite{Dehesa2021}). This is especially true for the Fisher-Shannon and LMC complexities of HOPs. However, recently, the Fisher-Shannon and LMC measures of the Hermite $H_n(x)$, Laguerre $L_n^{(\alpha)}(x)$ and Gegenbauer polynomials $C_n^{(\alpha)}(x)$ have been determined \cite{Sobrino2021} when $n\rightarrow \infty$ and when $\alpha\rightarrow \infty$, obtaining simple and transparent expressions.\\

 The goal of the present work is the analytical evaluation of the Cram\'er-Rao, Fisher-Shannon and LMC complexities of the whole family of Jacobi polynomials $\hat{P}_n^{(\alpha,\beta)}(x)$, with $\alpha,\beta>-1$, in the extreme situations ($n\rightarrow \infty$; fixed $\alpha,\beta$) and  ($\alpha\rightarrow \infty$; fixed $n,\beta$). These polynomials \cite{Jacobi1826,Jacobi1859,Szego1939,Nikiforov1988,Olver2010,Yang2021} are known to be orthonormal with respect to the weight function $h_{\alpha,\beta}(x) =(1-x)^{\alpha} (1+x)^{\beta}$ as 
   \begin{equation}\label{eq:ortho}
   	\int_{-1}^{+1} \hat{P}_n^{(\alpha,\beta)}(x) \hat{P}_m^{(\alpha,\beta)}(x)\, h_{\alpha,\beta}(x)\, dx = \delta_{mn}.
   \end{equation}  
Then, the associated Rakhmanov probability density $\rho_n(x)$ is given by
 \begin{equation}\label{eq:rakhmanov}
 	 \rho_n(x) =\left[\hat{P}_n^{(\alpha,\beta)}(x)\right]^2 \,h_{\alpha,\beta}(x).
 \end{equation} 
Moreover we will denote the orthogonal Jacobi polynomials  $P_{n}^{(\alpha,\beta)}(x)= \hat{P}_{n}^{(\alpha,\beta)}(x)\,(\kappa_n)^{1/2}$, with the normalization constant given (see e.g. \cite{Olver2010}) by
\begin{equation}
	\kappa_n = \int_{-1}^{+1} |P_n^{(\alpha,\beta)}(x)|^2\, h_{\alpha,\beta}(x) dx = \frac{2^{\alpha+\beta+1}\Gamma(\alpha+n+1)
\Gamma(\beta+n+1)}{n!(\alpha+\beta+2n+1)\Gamma(\alpha+\beta+n+1)}.
\end{equation}
The special case $\alpha=\beta=\lambda-\frac{1}{2}$ corresponds to the ultraspherical or Gegenbauer polynomials $C_n^{(\lambda)}(x),\lambda> -\frac{1}{2},$ with slightly different normalization, and the case $\alpha=\beta=0$ corresponds to the Legendre polynomials (see e.g. \cite{Olver2010}).\\

The structure of this paper is as follows.  In Sections \ref{CramerRao}, \ref{FisherShannon} and \ref{LMC} we obtain the asymptotic behavior for the Cram\'er-Rao, Fisher-Shannon and LMC complexities of the Jacobi polynomials $P_n^{(\alpha,\beta)}(x)$ when ($n\rightarrow \infty; \mbox{fixed}\, \alpha,\beta$) and when ($\alpha\rightarrow \infty; \mbox{fixed}\, n, \beta$) in a simple, compact and transparent form, respectively. Then, some concluding remarks and a few open related issues are pointed out.

\section{Cram\'er-Rao complexity of Jacobi polynomials}
\label{CramerRao}

The Cram\'er-Rao complexity of the Jacobi polynomials is given by the corresponding quantity of its associated Rakhmanov density (\ref{eq:rakhmanov}), which quantifies the combined balance of the pointwise probability concentration over its support interval jointly with the spreading of the probability around the centroid. It is defined \cite{Dembo1991,Dehesa2006,Antolin2009} by
\begin{equation}
      \label{eq:cramerrao}
      \mathcal{C}_{CR}[\hat{P}_n^{(\alpha,\beta)}]=F[\hat{P}_n^{(\alpha,\beta)}] \times V[\hat{P}_n^{(\alpha,\beta)}],
  \end{equation}
where $F[\hat{P}_n^{(\alpha,\beta)}]$ and $V[\hat{P}_n^{(\alpha,\beta)}]$ are the Fisher information \cite{Fisher1925,Frieden2004} and the variance of the Rakhmanov density (\ref{eq:rakhmanov}), which are defined as
 \begin{equation*}
     F\left[\hat{P}_n^{(\alpha,\beta)}\right]=\int_{-1}^{+1} \frac{[\rho'_n(x)]^2}{\rho_n(x)}dx,\quad \mbox{and}\quad V[\hat{P}_n^{(\alpha,\beta)}]=\langle x^2 \rangle-\langle x \rangle^2,
  \end{equation*}
respectively, with the expectation value $\langle x^k \rangle= \int_{-1}^{+1} x^k \rho_n(x) dx$ for $k=1,2$. In this section we give the explicit expressions of these three spreading quantities and we find in a simple compact form the values of the Cram\'er-Rao complexity of the Jacobi polynomials in the two following asymptotical regimes:  ($n\rightarrow \infty$; fixed $\alpha,\beta$) and  ($\alpha\rightarrow \infty$; fixed $n,\beta$). 

The particularly elegant algebraic properties of the Jacobi polynomials (see e.g. \cite{Nikiforov1988,Olver2010,Yang2021}) have allowed to encounter the following expression
\begin{eqnarray}
       \label{fisherjacobi}
        F\left[\hat{P}_n^{(\alpha,\beta)}\right]=
       \left\{
       \begin{array}{ll}
       2n(n+1)(2n+1), & \alpha,\beta=0, \\[3mm]
       \frac{2n+\beta+1}{4}\left[\frac{n^2}{\beta+1}+n+(4n+1)(n+\beta+1)+\frac{(n+1)^2}{\beta-1}\right],
       & \alpha=0, \beta>1, \\[3mm]
       \frac{2n+\alpha+\beta+1}{4(n+\alpha+\beta-1)}\left[n(n+\alpha+\beta-1)
       \left(\frac{n+\alpha}{\beta+1}+2+\frac{n+\beta}{\alpha+1}\right)\right.& \\
       +\left.(n+1)(n+\alpha+\beta)\left(\frac{n+\alpha}{\beta-1}+2+\frac{n+\beta}{\alpha-1}\right)\right],
       &\alpha, \beta>1, 
       \end{array}
       \right.
    \end{eqnarray}
    (and $\infty$ otherwise) for the Fisher information \cite{SanchezRuiz2005,Guerrero2010}, and 
   \begin{eqnarray}
   \label{eq:variance_jacobi}
     V[\hat{P}_n^{(\alpha,\beta)}]=
     \frac{4(n+1)(n+\alpha+1)(n+\beta+1)(n+\alpha+\beta+1)}{(2n+\alpha+\beta+1)(2n+\alpha+\beta+2)^2 (2n+\alpha+\beta+3)}\nonumber\\
     +\frac{4n(n+\alpha)(n+\beta)(n+\alpha+\beta)}{(2n+\alpha+\beta-1)(2n+\alpha+\beta)^2 (2n+\alpha+\beta+1)},
     \label{eq:variance}
   \end{eqnarray}
   for the variance \cite{Dehesa2006} of Jacobi polynomials. Then, from 
Eqs.(\ref{eq:cramerrao})-(\ref{eq:variance}) one has \cite{Dehesa2015} the following values for the Cram\'er-Rao complexity of the Jacobi polynomials  
    \begin{eqnarray*}
       C_{CR}\left[\hat{P}_n^{(\alpha,\beta)}\right]=
      \left\{
      \begin{array}{ll}
      2n(n+1)\left[\frac{(n+1)^2}{2n+3}+\frac{n^2}{2n-1}\right],& \alpha=\beta=0, \\ & \\
      \left[\frac{(n+1)^2(n+\beta+1)^2}{(2n+\beta+2)^2(2n+\beta+3)}+\frac{n^2(n+\beta)^2}{(2n+\beta-1)(2n+\beta)^2}\right] & \\
      \times \left[\frac{n^2}{\beta+1}+n+(4n+1)(n+\beta+1)+\frac{(n+1)^2}{\beta-1}\right], & \alpha=0,\beta>1, \\ & \\
      \left[\frac{(n+1)(n+\alpha+1)(n+\beta+1)(n+\alpha+\beta+1)}{(2n+\alpha+\beta+2)^2(2n+\alpha+\beta+3)}+
      \frac{n(n+\alpha)(n+\beta)(n+\alpha+\beta)}{(2n+\alpha+\beta-1)(2n+\alpha+\beta)^2}\right] & \\
      \times \frac{1}{n+\alpha+\beta-1}\left[n(n+\alpha+\beta-1)\left(\frac{n+\alpha}{\beta+1}+2+\frac{n+\beta}{\alpha+1}\right)\right. & \\
      \left.+(n+1)(n+\alpha+\beta)\left(\frac{n+\alpha}{\beta-1}+2+\frac{n+\beta}{\alpha-1}\right)\right], & \alpha>1,\beta>1,       \end{array}
      \right.
     \end{eqnarray*}
 From this expression we easily obtain the high-degree asymptotics ($n\rightarrow \infty$; fixed $\alpha,\beta$) 
     \begin{eqnarray}
       C_{CR}\left[\hat{P}_n^{(\alpha,\beta)}\right]=
      \left\{
      \begin{array}{ll}
      2n^{3}+\mathcal{O}(n^{2}),& \alpha=\beta=0, \\ & \\
      \frac{1}{2}\left(2+\frac{\beta}{\beta^{2}-1}\right)n^{3}+\mathcal{O}(n^{2}), & \alpha=0,\beta>1, \\ & \\
      \frac{1}{2}\left(\frac{\alpha}{\alpha^{2}-1}+\frac{\beta}{\beta^{2}-1}\right)n^{3}+\mathcal{O}(n^{2}),& \alpha>1,\beta>1,       \end{array}
      \right.
     \end{eqnarray}
and the high-parameter asymptotics ($\alpha\rightarrow \infty$; fixed $n,\beta$) 
      \begin{align}
       C_{CR}\left[\hat{P}_n^{(\alpha,\beta)}\right]=
      \frac{(1+\beta+2n\beta)(1+\beta+2n(1+n+\beta))}{\beta^{2}-1},\qquad \beta>1,\qquad \alpha\to\infty,      
     \end{align}
for the Cram\'er-Rao complexity of the Jacobi polynomials. Summarizing, we first  observe that in the limit $n\rightarrow \infty$ the Cram\'er-Rao complexity of the Jacobi polynomials follow a qualitative $n^3$-law, similarly to the corresponding quantity of Laguerre polynomials \cite{Dehesa2015}, despite the fact that the respective weight functions are so different. This is because the two factors (variance and Fisher information) of the Cram\'er-Rao complexity of Laguerre polynomials have a linear (Fisher infomation) and quadratic (variance) dependence on $n$, while for the Jacobi polynomials the Fisher infomation has a cubic dependence on the degree and the variance is constant. Moreover, in the limit ($\alpha\rightarrow \infty$; fixed $n,\beta$) the Cram\'er-Rao complexity of Laguerre and Jacobi polynomials have also a mutual similar behavior, having a constant leading term that depends on $n,\beta$ for the Jacobi polynomals and $n$ for the Laguerre polynomials. In this limit the Fisher information and variance have a direct and inverse quadratic dependence on $\alpha$ for the Jacobi polynomials, while for the Laguerre polynomials the Fisher information and the variance have an inverse and direct linear dependence on $\alpha$, respectively.

\section{Fisher-Shannon complexity of Jacobi polynomials}
\label{FisherShannon}

This statistical quantity is given by the Fisher-Shannon complexity of the Rakhmanov density (\ref{eq:rakhmanov}) of the Jacobi polynomials, which is defined \cite{Angulo2008,Romera2004} as 
\begin{eqnarray}
     \label{fishershannon}
      \mathcal{C}_{FS}[\hat{P}_n^{(\alpha,\beta)}]=F[\hat{P}_n^{(\alpha,\beta)}] \times \frac{1}{2 \pi e} e^{2 S[\hat{P}_n^{(\alpha,\beta)}]}=\frac{1}{2 \pi e} F[\hat{P}_n^{(\alpha,\beta)}] \times \left(\mathcal{L}_{S}[\hat{P}_n^{(\alpha,\beta)}] \right)^2,
  \end{eqnarray}
  where the symbols $F[\hat{P}_n^{(\alpha,\beta)}]$ and  $\mathcal{L}_{S}[p_n]=e^{S[p_n]}$ denotes the Fisher information and the Shannon entropic power or Shannon spreading length of the polynomial $\hat{P}_n^{(\alpha,\beta)}(x)$, respectively. Note that $\mathcal{C}_{FS}[\hat{P}_n^{(\alpha,\beta)}]$ measures the gradient content of the Rakhmanov probability density $\rho_n(x)$ associated to the polynomial $\hat{P}_n^{(\alpha,\beta)}(x)$ and its total extent along the support interval $[-1,+1]$ simultaneously.\\

  The explicit expression of the Fisher-Shannon complexity of the Jacobi polynomials for generic values $(n,\,\alpha,\beta)$ is unknown up until now. This is so because, although the Fisher information has been given by Eq. (\ref{fisherjacobi}), the Shannon entropy is not known in spite of many efforts. However, there are two extreme situations in which the value of this quantity can be analytically evaluated; namely, when ($\alpha \rightarrow \infty; \mbox{fixed}\, n, \beta$) and when ($n\rightarrow \infty; \mbox{fixed}\, \alpha,\beta$). The goal of this section is to obtain these two parameter and degree asymptotics in a compact way for the Shannon spreading length, and then for the Fisher-Shannon complexity (Eq. \ref{fishershannon}) of the orthonormal Jacobi polynomials $\hat{P}_n^{(\alpha,\beta)}(x)$. The main results are  briefly summarized in Table  \ref{Table_1}.\\
  
\begin{table}
\centering
\begin{tabular}{|c|c|c|}
\hline
Measure  of $\hat{P}_n^{(\alpha,\beta)}(x)$ & n$\to \infty$ & $\alpha\to \infty$ \\ \hline
$\mathcal{L}_{S}[\hat{P}_n^{(\alpha,\beta)}]$ & $\left.\frac{\pi}{e}\right.$ & $\frac{1}{\alpha}$ \\ \hline
$\mathcal{C}_{FS}[\hat{P}_n^{(\alpha,\beta)}]$ & $\begin{array}{ll}
\frac{2\pi}{e^{3}}n^{3}& \alpha,\beta=0 \\
\frac{\pi}{4e^{3}}\left(4+\frac{1}{\beta-1}+\frac{1}{\beta+1}\right)n^{3}
& \alpha=0, \beta>1 \\
\frac{\pi(\alpha+\beta)(\alpha\beta-1)}{2e^{3}(\alpha^{2}-1)(\beta^{2}-1)}n^{3}
&\alpha, \beta>1 \\
\infty & {\rm otherwise}
\end{array}$
 & $ \frac{(1+\beta+2n\beta)}{8\pi e(\beta^{2}-1)} $ \\ \hline
\end{tabular}
\caption{First order asymptotics for the  Shannon spreading length $\mathcal{L}_{S}$ and the Fisher-Shannon complexity $\mathcal{C}_{FS}$  measures of the orthonormal Jacobi polynomials $\hat{P}_n^{(\alpha,\beta)}(x)$, when $n\to \infty$ and $\alpha\to \infty$. }
\label{Table_1}
\end{table}

First, we realize that the Shannon-like integral functional of the orthonormal Jacobi polynomials $\hat{P}_n^{(\alpha,\beta)}(x)$ is given by
\begin{align}\label{shannonJac}
S\left[\hat{P}_n^{(\alpha,\beta)}\right]=&-\int_{-1}^{+1}
\left[\hat{P}_n^{(\alpha,\beta)}(x)\right]^2h_{\alpha,\beta}(x) \log\left\{
\left[\hat{P}_n^{(\alpha,\beta)}(x)\right]^2 h_{\alpha,\beta}(x) \right\}dx
=E\left[\hat{P}_n^{(\alpha,\beta)}\right]+I\left[\hat{P}_n^{(\alpha,\beta)}\right],
\end{align}
with the functional \cite{Sanchez2000}
\begin{align} \label{IntegralGeg0}
	I\left[\hat{P}_n^{(\alpha,\beta)}\right] =& -\int_{-1}^{+1}
\left[\hat{P}_n^{(\alpha,\beta)}(x)\right]^2 h_{\alpha,\beta}(x) \log h_{\alpha,\beta}(x) \nonumber\\
 =& (\alpha+\beta)\left(\frac{1}{2n+\alpha+\beta+1}+2\psi(2n+\alpha+\beta+1)-\psi(n+\alpha+\beta+1)-\log(2)\right)\nonumber\\
&-\left( \alpha\psi(n+\alpha+1)+\beta\psi(n+\beta+1)\right)
 \end{align}
and the Shannon entropy $E\left[\hat{P}_n^{(\alpha,\beta)}\right]$, which is  defined by
\begin{align}\label{eq_entropic_functional}
E\left[\hat{P}_n^{(\alpha,\beta)}\right]&=-\int_{-1}^{+1}
\left[\hat{P}_n^{(\alpha,\beta)}(x)\right]^2 h_{\alpha,\beta}(x) \log
\left[\hat{P}_n^{(\alpha,\beta)}(x)\right]^2dx .
\end{align}
The analytical determination of this entropic measure is a formidable task. Indeed, it has have been  calculated for integer values of the polynomial's parameter in a somewhat highbrow manner \textit{only}. However, we find below that they can be expressed in a simple and compact way for the two extreme situations mentioned above. 
  \subsection{Asymptotics $n\rightarrow \infty$}
  
The Shannon entropy of the Jacobi polynomials has been shown to have the following degree asymptotics 
\begin{equation}\label{EJac}
	E({\hat{P}_n^{(\alpha,\beta)}})=  \log(\pi)-1-(\alpha+\beta)\log(2)+\mathcal{O}(n^{-1}),\qquad n\rightarrow \infty
\end{equation}
for fixed $(\alpha, \beta)$ \cite{Aptekarev1994,Dehesa2001}. Moreover, from Eq. (\ref{IntegralGeg0}) and the known asymptotical behavior \cite{Olver2010} of the involved gamma $\Gamma(x)$ and digamma $\psi(x)$ functions, we find that for fixed $(\alpha,\beta)$ the integral functional $I\left[\hat{P}_n^{(\alpha,\beta)}\right]$ fulfills the asymptotics 
\begin{align}\label{IJacinf}
I\left[\hat{P}_n^{(\alpha,\beta)}\right] =(\alpha+\beta)\log(2)+\mathcal{O}(n^{-1}),\qquad n\rightarrow \infty
\end{align}
Therefore, from Eqs. (\ref{shannonJac}), (\ref{EJac}) and (\ref{IJacinf}) we find that the asymptotics for the Shanon-like functional of the Jacobi polynomials is
\begin{align}
S\left[\hat{P}_n^{(\alpha,\beta)}\right] =  \log(\pi)-1+\mathcal{O}(n^{-1}),\qquad n\rightarrow \infty
\end{align}
so that the Shannon spreading length of the Jacobi polynopmials has the behavior
\begin{equation}\label{ANLSG}
	\mathcal{L}_{S}[\hat{P}_n^{(\alpha,\beta)}] \sim \frac{\pi}{e} , \qquad n\rightarrow \infty
\end{equation}
On the other hand, from expression (\ref{fisherjacobi}) we can obtain the following ($n\rightarrow \infty; \mbox{fixed}\, \alpha,\beta$) asymptotics for the Fisher information of the Jacobi polynomials :
\begin{align}\label{ANFSG}
F\left[\hat{P}_n^{(\alpha,\beta)}\right]=
\left\{
\begin{array}{ll}
4n^{3}+\mathcal{O}(n^2), & \alpha,\beta=0, \\
\frac{1}{2}\left(4+\frac{1}{\beta-1}+\frac{1}{\beta+1}\right)n^{3}+\mathcal{O}(n^2),
& \alpha=0, \beta>1, \\
\frac{(\alpha+\beta)(\alpha\beta-1)}{(\alpha^{2}-1)(\beta^{2}-1)}n^{3}+\mathcal{O}(n^2),
&\alpha, \beta>1, \\
\infty, & {\rm otherwise.}
\end{array}
\right.
\end{align}

Finally, taking into account (\ref{fishershannon}), (\ref{ANLSG}) and (\ref{ANFSG}),  we have that the Fisher-Shannon complexity of the Jacobi polynomials has the following asymptotics ($n\rightarrow \infty; \mbox{fixed}\, \alpha,\beta$) behavior
 \begin{align}\label{FSJAC}
      \mathcal{C}_{FS}\left[\hat{P}_n^{(\alpha,\beta)}\right]=
      \left\{
\begin{array}{ll}
\frac{2\pi}{e^{3}}n^{3}+\mathcal{O}(n^2), & \alpha,\beta=0, \\
\frac{\pi}{4e^{3}}\left(4+\frac{1}{\beta-1}+\frac{1}{\beta+1}\right)n^{3}+\mathcal{O}(n^2),
& \alpha=0, \beta>1, \\
\frac{\pi(\alpha+\beta)(\alpha\beta-1)}{2e^{3}(\alpha^{2}-1)(\beta^{2}-1)}n^{3}+\mathcal{O}(n^2),
&\alpha, \beta>1, \\
\infty, & {\rm otherwise.}
\end{array}
\right.
\end{align}
which extends and includes the corresponding asymptotical quantity recently obtained for the subfamily of Gegenbauer polynomials \cite{Dehesa2021} to the whole Jacobi family of orthogonal polynomials.  Moreover, we observe that in the limit $n\rightarrow \infty$ the Fisher-Shannon complexity of the Jacobi polynomials behaves qualitatively similar to the corresponding quantity of Laguerre polynomials \cite{Sobrino2021}, following a $n^3$-law, despite the fact that the weight functions are very different in each case; this is because for the Laguerre polynomials both the Fisher information and the Shannon spreading length have a linear dependence on $n$, while for the Jacobi polynomials the Fisher information has a cubic dependence on the degree and the Shannon spreading length is constant. 

 Interestingly, for the quantum systems with a solvable quantum-mechanical potential with bound-states wavefunctions controlled by Jacobi polynomials (e.g., some supersymmetric quantum systems) (see e.g. \cite{Ghosh2019,Chaitanya2012,Cooper2001,Gangopadhyaya2017}),
 the Fisher-Shannon measure (\ref{FSJAC}) allows one to find the quantum-classical limit of the physical Fisher-Shannon complexity; they correspond to the high-energy or Rydberg states  since for such a limit $n\rightarrow \infty$ the wavelengths of particles are small in comparison with the characteristic dimensions of the system and the wavefunctions of the quasi-classical state.
%
%

\subsection{Asymptotics $\alpha \rightarrow \infty$}
\label{asympspreading}

Now we determine the Fisher-Shannon complexity $\mathcal{C}_{FS}[\hat{P}_n^{(\alpha,\beta)}]$, given by (\ref{fishershannon}), in the limit $\alpha \rightarrow \infty$ with fixed degree $n, \beta$. We start by evaluating the Shannon entropy (\ref{eq_entropic_functional}) of the orthogonal Jacobi polynomials in this limit by means of the relation
\begin{equation}\label{EPN2}
	E\left[P_n^{(\alpha,\beta)}\right] =2 \frac{d}{dp}\left[\mathcal{N}_{p}\left[P_n^{(\alpha,\beta)}\right]\right]_{p=2},
\end{equation}
where the symbol $\mathcal{N}_{p}$ denotes the norm  of the orthogonal Jacobi polynomials defined as
\begin{align}\label{Np_norm}
\mathcal{N}_{p}\left[P_n^{(\alpha,\beta)}\right]=\int_{-1}^{1}(1-x)^\alpha (1+x)^\beta\left|P_n^{(\alpha,\beta)}(x)\right|^{p}dx.
\end{align}
This quantity can be analytically estimated for $\alpha \to\infty$ by taking into account the known relation \cite{Olver2010}
\begin{equation}\label{limitJac}
	\lim_{\alpha \rightarrow \infty} \frac{P_n^{(\alpha,\beta)}(x)}{P_n^{(\alpha,\beta)}(1)} = \left(\frac{1+x}{2}\right)^{n}, \qquad \text{with}\qquad P_n^{(\alpha,\beta)}(1) = \frac{\Gamma(\alpha+n+1)}{n!\,\Gamma(\alpha+1)}.
\end{equation}
Then, from (\ref{Np_norm}) and (\ref{limitJac}) we have the asymptotics
\begin{align}\label{Np_norm_2}
\mathcal{N}_{p}\left[P_n^{(\alpha,\beta)}\right]&\sim  P_n^{(\alpha,\beta)}(1)\,2^{-np}\left(\frac{_{2}F_{1}(1,-\alpha,2+np+\beta;-1)}{1+np+\beta}+\frac{_{2}F_{1}(1,-np-\beta,2+\alpha;-1)}{1+\alpha}\right)\nonumber\\
&= P_n^{(\alpha,\beta)}(1)\,2^{-np}(1+np+\beta)^{-1}(1+\alpha)^{-1}\times\nonumber\\
&\left((1+\alpha){}\,_{2}F_{1}(1,-\alpha,2+np+\beta;-1)+(1+np+\beta){}\,_{2}F_{1}(1,-np-\beta,2+\alpha;-1) \right),
\end{align}
where $_{2}F_{1}(a,b,c;x)$ denotes the Gaussian hypergeometric function \cite{Olver2010}. Then, 
taking into account the known relation between the hypergeometric functions
\begin{align}
(1-a){}_{2}F_{1}(1,a,2-b,-1)+(1-b){}_{2}F_{1}(1,b,2-a,-1)=\frac{2^{1-a-b}\Gamma(2-a)\Gamma(2-b)}{\Gamma(2-a-b)},
\end{align}
with the parameters $a=-\alpha$ and $b=-np-\beta$, the asymptotical behavior (\ref{Np_norm_2}) simplifies as
\begin{align}\label{Np_norm_2_simplified}
\mathcal{N}_{p}\left[P_n^{(\alpha,\beta)}\right]&\sim   \frac{\Gamma(\alpha+n+1)}{n!}
\frac{\Gamma(1+np+\beta)}{\Gamma(2+\alpha+np+\beta)}2^{1+\alpha+\beta}.
\end{align}
Thus, according to Eqs. (\ref{EPN2}) and (\ref{Np_norm_2_simplified}), one has that the Shannon entropy  of the orthogonal Jacobi polynomials $P_n^{(\alpha,\beta)}(x)$ in the current limit is given as
\begin{align}
	E\left[P_n^{(\alpha,\beta)}\right]&	\sim  2^{2+\alpha+\beta}\frac{\Gamma(1+n+\alpha)\Gamma(1+2n+\beta)}{\Gamma(n)\Gamma(2+2n+\alpha+\beta)}(\psi(1+2n+\beta)-\psi(2+2n+\alpha+\beta))\nonumber\\
	&=2^{2+\alpha+\beta}\alpha^{-n-\beta-1}\left(\frac{\Gamma(1+2n+\beta)}{\Gamma(n)}(\psi(1+2n+\beta)-\log(\alpha))+\mathcal{O}(\alpha^{-2})\right),\qquad  \alpha \rightarrow \infty,\nonumber
\end{align}
so that we can express the Shannon entropy  of  the orthonormal Jacobi polynomials as
\begin{align}
	E\left[\hat{P}_n^{(\alpha,\beta)}\right] = \frac{1}{\kappa^{P}_{n}}E\left[P_n^{(\alpha,\beta)}\right] +\log(\kappa_{n})=&2\alpha^{-n}\left(\frac{n\Gamma(1+2n+\beta)}{\Gamma(1+n+\beta)}(\psi(1+2n+\beta)-\log(\alpha))+\mathcal{O}(\alpha^{-2})\right)\nonumber\\&+(1+\alpha+\beta)\log(2)+\log(\frac{\Gamma(1+n+\beta)}{n!})-(1+\beta)\log(\alpha).
	\label{E_lambda_infinity}
\end{align}
Moreover, from (\ref{IntegralGeg0}), we have the following asymptotics for the auxiliary functional $I\left[P_n^{(\alpha,\beta)}\right]$ 
\begin{align}\label{ECn}
I\left[\hat{P}_n^{(\alpha,\beta)}\right]=-\alpha\log(2)+1+2n+\beta-\beta\log(2)+\beta\log(\alpha)-\beta\,\psi(1+n+\beta)+\mathcal{O}(\alpha^{-1}),\qquad \alpha \rightarrow \infty.
\end{align}
A similar result follows for $\beta\to\infty$ by exchanging $\alpha\leftrightarrow\beta$. 
Then, according to Eqs. (\ref{shannonJac}), (\ref{E_lambda_infinity}) and (\ref{ECn}), we find the following asymptotics for the Shannon-like integral functional of the Jacobi polynomials
\begin{align}
\label{SC_lambda_infinity}
S\left[\hat{P}_n^{(\alpha,\beta)}\right]\sim -\log(\alpha)+\mathcal{O}(1),\qquad  \alpha \rightarrow \infty,
\end{align}
so that the Shannon entropy power or spreading length of Jacobi polynomials behaves as 
\begin{equation}\label{LSCn}
	\mathcal{L}_{S}[\hat{P}_n^{(\alpha,\beta)}] \sim \frac{1}{\alpha}, \qquad \alpha \rightarrow \infty.
\end{equation}
On the other hand, the  asymptotics ($\alpha \rightarrow \infty$, fixed $n,\beta$) for the Fisher information (\ref{fisherjacobi}) of the Jacobi polynomials F[$\hat{P}_n^{(\alpha,\beta)}$] turns out to be
\begin{equation}
        \label{FC_lambda_infinity}
	F[\hat{P}_n^{(\alpha,\beta)}] = \frac{(1+\beta+2n\beta)}{4(\beta^{2}-1)}\alpha^{2}+\mathcal{O}(\alpha) ,\qquad \alpha \rightarrow \infty.
\end{equation}

Finally, the substitution of the last two quantities into Eq. (\ref{fishershannon}) gives rise to the following ($\alpha \rightarrow \infty$, fixed $n,\beta$)-asymptotics for the Fisher-Shannon complexity of the orthonormal Jacobi polynomials:
\begin{equation}
        \label{CFSJAC}
	\mathcal{C}_{FS}[\hat{P}_n^{(\alpha,\beta)}] \sim \frac{(1+\beta+2n\beta)}{8\pi e(\beta^{2}-1)}+\mathcal{O}(\alpha^{-1}) ,\qquad \alpha \rightarrow \infty.
\end{equation}
The corresponding result for the polynomials with ($\alpha\rightarrow \infty; \beta \rightarrow \infty; \mbox{fixed}\, n$) remains to be found; we have not been able to find it because the involved asymptotical behavior of the second-order entropic moment $W_2[\hat{P}_n^{(\alpha,\beta)}]$ is a non-trivial task.

Finally, for multidimensional quantum systems with stationary states controlled by Jacobi polynomials (e.g., some supersymmetric quantum systems) (see e.g. \cite{Ghosh2019,Chaitanya2012,Cooper2001,Gangopadhyaya2017}),
 the Fisher-Shannon measure (\ref{CFSJAC}) and its asymptotical extension for ($\alpha\rightarrow \infty; \beta \rightarrow \infty; \mbox{fixed}\, n$) allow us to find the pseudo-classical limit of the physical Fisher-Shannon complexity; they correspond to the high-dimensional or pseudoclassical states. The latter is because the wavefunctions of such extreme states involve polynomials orthogonal with respect to a Jacobi weight function where both parameters $\alpha$ and $\beta$ are directly proportional to the system's dimensionality.

\section{LMC complexity of Jacobi polynomials}
\label{LMC}

In this section we investigate the LMC complexity of the orthonormal Jacobi polynomials $\hat{P}_n^{(\alpha,\beta)}(x)$, which is defined as 
   \begin{equation}\label{lmcdefJac}
      \mathcal{C}_{LMC}[\hat{P}_n^{(\alpha,\beta)}] = W_2[\hat{P}_n^{(\alpha,\beta)}] \times \mathcal{L}_{S}[\hat{P}_n^{(\alpha,\beta)}],
   \end{equation}
   where the second-order entropic moment $W_2[\hat{P}_n^{(\alpha,\beta)}]$, which measures the disequilibrium or deviation from uniformity, is given by
   \begin{equation}\label{omega2Jac}
	W_2[\hat{P}_n^{(\alpha,\beta)}] = \int_{-1}^{+1}
\left(\left[\hat{P}_n^{(\alpha,\beta)}(x)\right]^2 h_{\alpha,\beta}(x)\right)^2\,dx = \int_{-1}^{+1} \,(1-x)^{2\alpha} (1+x)^{2\beta}\,\left[\hat{P}_n^{(\alpha,\beta)}(x)\right]^4\,dx.
\end{equation}
This statistical complexity quantifies the combined balance of the disequilibrium and the total extent of the polynomials along its weight function. The explicit expression of this measure in terms of the degree $n$ and the parameters $(\alpha, \beta)$ has not yet been determined in a handy way, because neither $W_2[\hat{P}_n^{(\alpha,\beta)}]$ nor the spreading length $\mathcal{L}_{S}[\hat{P}_n^{(\alpha,\beta)}]$ are analytically known. In this section we obtain simple and compact analytical expressions for  $\mathcal{C}_{LMC}[\hat{P}_n^{(\alpha,\beta)}]$ in the two extreme situations, ($n\rightarrow \infty; \mbox{fixed}\, \alpha,\beta$) and ($\alpha\rightarrow \infty; \mbox{fixed}\, n,\beta$). They are briefly summarized in Table  \ref{Table_2}. \\

   \begin{table}
\centering
\begin{tabular}{|c|c|c|}
\hline
Measure  of $\hat{P}_n^{(\alpha,\beta)}(x)$ & n$\to \infty$ & $\alpha\to \infty$ \\ \hline
$W_2[\hat{P}_n^{(\alpha,\beta)}]$ & $\begin{array}{ll}
\frac{2^{\alpha+\beta-2}3}{\pi^{2}}\frac{\Gamma(\alpha)\Gamma(\beta)}{\Gamma(\alpha+\beta)} & \beta>0 \\
\log{(n)}
& \beta=0 \\
n^{-2\beta} &-1<\beta<0
\end{array}$ & $\frac{\Gamma(1+4n+2\beta)}{2^{2(1+2n+\beta)}n!^{2}\Gamma(1+n+\beta)}\alpha$ \\ \hline
$\mathcal{C}_{LMC}[\hat{P}_n^{(\alpha,\beta)}]$ & $\begin{array}{ll}
\frac{2^{\alpha+\beta-2}3}{\pi e}\frac{\Gamma(\alpha)\Gamma(\beta)}{\Gamma(\alpha+\beta)} & \beta>0 \\
\frac{\pi}{e}\log{(n)}
& \beta=0 \\
\frac{\pi}{e}n^{-2\beta} &-1<\beta<0
\end{array}$
 & $ \frac{\Gamma(1+4n+2\beta)}{2^{2(1+2n+\beta)}n!^{2}\Gamma(1+n+\beta)}$ \\ \hline
\end{tabular}
\caption{First order asymptotics for the  disequilibrium $W_2$ and the LMC complexity $\mathcal{C}_{LMC}$  measures of the orthonormal Jacobi polynomials $\hat{P}_n^{(\alpha,\beta)}(x)$, when $n\to \infty$ and $\alpha\to \infty$. }
\label{Table_2}
\end{table}

 \subsection{Asymptotics $n\rightarrow \infty$}
 
 To determine the ($n\rightarrow \infty; \mbox{fixed}\, \alpha,\beta$)-asymptotics of the LMC complexity $\mathcal{C}_{LMC}[\hat{P}_n^{(\alpha,\beta)}]$ we start by calculating the disequilibrium $W_2[\hat{P}_n^{(\alpha,\beta)}]$ in the limit $n\rightarrow \infty$ by means of Theorem 3 of Aptekarev et al \cite{Aptekarev2021}, obtaining
\begin{align}\label{omegaGeg_n_infinity}
	W_2[\hat{P}_n^{(\alpha,\beta)}]\sim
	\left\{
\begin{array}{ll}
\frac{2^{\alpha+\beta-2}3}{\pi^{2}}\frac{\Gamma(\alpha)\Gamma(\beta)}{\Gamma(\alpha+\beta)}, & \beta>0, \\
\log{(n)},
& \beta=0, \\
n^{-2\beta}, &-1<\beta<0,
\end{array}
\right.
\end{align}
and then we keep in mind the corresponding asymptotics (\ref{ANLSG}) for the Shannon spreading length $\mathcal{L}_{S}[\hat{P}_n^{(\alpha,\beta)}]$. The substitution of the asymptotical values of these two entropic quantities into Eq. (\ref{lmcdefJac}) gives rise to the following asymptotical behavior ($n\rightarrow \infty$) of the LMC complexity of the orthonormal Jacobi polynomials $\hat{P}_n^{(\alpha,\beta)}(x)$:
 \begin{align}\label{lmcJac_n_infinity}
      \mathcal{C}_{LMC}[\hat{P}_n^{(\alpha,\beta)}] \sim     	\left\{
\begin{array}{ll}
\frac{2^{\alpha+\beta-2}3}{\pi e}\frac{\Gamma(\alpha)\Gamma(\beta)}{\Gamma(\alpha+\beta)}, & \beta>0, \\
\frac{\pi}{e}\log{(n)},
& \beta=0, \\
\frac{\pi}{e}n^{-2\beta}, &-1<\beta<0,
\end{array}
\right.
\end{align}
\begin{figure}
	\centering
	\includegraphics[width=0.7\linewidth]{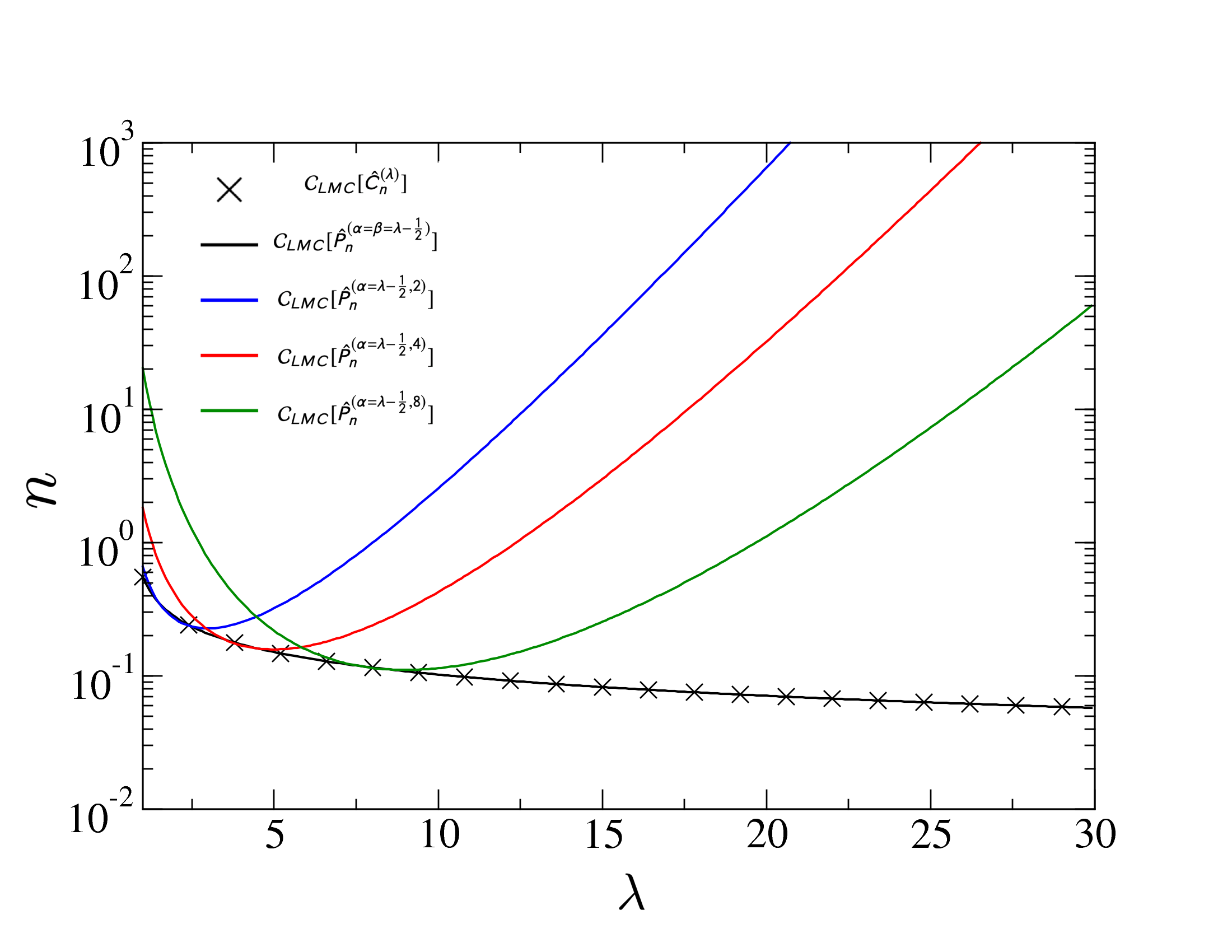}
	\caption{Comparison of the LMC complexity $\mathcal{C}_{LMC}$  measures of the orthonormal Gegenbauer polynomials $\hat{C}_n^{(\lambda)}(x)$ and the orthonormal Jacobi polynomials $\hat{P}_n^{(\alpha,\beta)}(x)$, with $(\alpha=\lambda-2$ and $\beta=\lambda-2, 2, 4, 8)$, for various values of $\lambda$ and $n \rightarrow \infty$.}
	\label{fig:lmc}
\end{figure}
which extends and includes the corresponding asymptotical quantity recently obtained for the subfamily of Gegenbauer polynomials \cite{Dehesa2021} to the whole Jacobi family of orthogonal polynomials. See also Figure 1, where the LMC complexity $\mathcal{C}_{LMC}$  measures of the orthonormal Jacobi polynomials $\hat{P}_n^{(\alpha,\beta)}(x)$, with $(\alpha=\lambda-2$ and $\beta=\lambda-2, 2, 4, 8)$, and the orthonormal Gegenbauer polynomials $\hat{C}_n^{(\lambda)}(x)$ are compared for various values of $\lambda$ and $n \rightarrow \infty$. Note that both Jacobi and Gegenbauer complexities match when $\beta=\lambda-2$ as one would expect, what is a partial checking of our results. On the other hand, we observe that in the limit $n\rightarrow \infty$ the LMC complexity of the Jacobi polynomials behaves very different to the corresponding quantity of Laguerre polynomials \cite{Dehesa2021}, as one expects because of their weight functions are so distinct, except in the special case $\beta=0$.  Indeed, for $\beta=0$ there happens the following phenomenon: the disequilibrium has a logarithmic dependence on $n$ while the Shannon spreading length is constant so that the total balance for the LMC complexity of the Jacobi polynomials obeys the $\log n$-law  as also occurs for the LMC complexity of Laguerre polynomials \cite{Dehesa2021}.\\

Here again, for the quantum systems with a solvable quantum-mechanical potential with bound-states wavefunctions controlled by Jacobi polynomials (e.g., some supersymmetric quantum systems) (see e.g. \cite{Ghosh2019,Chaitanya2012,Cooper2001,Gangopadhyaya2017}),
 the LMC measure (\ref{FSJAC}) allows one to find the quantum-classical limit of the physical LMC complexity; they correspond to the high-energy or Rydberg states.

 \subsection{Asymptotics $\alpha\rightarrow \infty$}

 Here we determine the LMC complexity $\mathcal{C}_{LMC}[\hat{P}_n^{(\alpha,\beta)}]$, given by Eq. (\ref{lmcdefJac}),  in the limit ($\alpha\rightarrow \infty; \mbox{fixed}\, n,\beta$).  First we realize that the Shannon spreading length $\mathcal{L}_{S}[\hat{P}_n^{(\alpha,\beta)}]$, has been already obtained in Eq. (\ref{LSCn}). Then, we calculate the second-order entropic moment $W_2[\hat{P}_n^{(\alpha,\beta)}]$, given by Eq. (\ref{omega2Jac}). We use the limiting relation (\ref{limitJac}) into (\ref{omega2Jac}), obtaining the value
\begin{align}
	W_2[P_n^{(\alpha,\beta)}]\sim& P_n^{(\alpha,\beta)}(1) ^{4}2^{-4n}\Gamma(1+2\alpha)\Gamma(1+4n+2\beta)\times\nonumber\\ &\left( \frac{1}{\Gamma(1+2\alpha)}{}_{2}\tilde{F}_{1}(1,-2\alpha,2+4n+2\beta,-1)+\frac{1}{\Gamma(1+4n+2\beta)} {}_{2}\tilde{F}_{1}(1,-4n-2\beta,2+2\alpha,-1) \right)\nonumber\\
	=&\frac{\Gamma(\alpha+n+1)^{4}}{(n!)^{4}\Gamma(\alpha+1)^{4}}\frac{\Gamma(1+4n+2\beta)\Gamma(1+2\alpha)}{\Gamma(2+2\alpha+4n+2\beta)}2^{1+2\alpha+2\beta}
\end{align}
for the orthogonal Jacobi polynomials. Now, the corresponding asymptotics for the second-order entropic power of the orthonormal Jacobi polynomials is given by
\begin{equation}\label{W2oG}
	W_2[\hat{P}_n^{(\alpha,\beta)}] = \frac{1}{(\kappa_{n})^2}\,W_2[P_n^{(\alpha,\beta)}] \sim \frac{\Gamma(1+4n+2\beta)}{2^{2(1+2n+\beta)}\,(n!)^{2}\,\Gamma(1+n+\beta)}\alpha ,\qquad \alpha\rightarrow \infty.
\end{equation}
Finally, the combination of Eqs. (\ref{lmcdefJac}), (\ref{LSCn}) and (\ref{W2oG}) lead to the asymptotical behavior
\begin{equation}\label{LMCJAC}
      \mathcal{C}_{LMC}[\hat{P}_n^{(\alpha,\beta)}] = \frac{\Gamma(1+4n+2\beta)}{2^{2(1+2n+\beta)}(n!)^{2}\,\Gamma(1+n+\beta)}, \qquad \alpha \rightarrow \infty 
   \end{equation}
for the LMC complexity of the (orthonormal) Jacobi polynomials with ($\alpha\rightarrow \infty; \mbox{fixed}\, n,\beta$). It remains to find the corresponding result for the polynomials with ($\alpha\rightarrow \infty; \beta \rightarrow \infty; \mbox{fixed}\, n$), which we have not been able to find because the involved asymptotical behavior of the the second-order entropic moment $W_2[\hat{P}_n^{(\alpha,\beta)}]$ is a non-trivial task.

Finally, for multidimensional quantum systems with bound states controlled by Jacobi polynomials (e.g., some supersymmetric quantum systems) (see e.g. \cite{Ghosh2019,Chaitanya2012,Cooper2001,Gangopadhyaya2017}),
 the LMC measure (\ref{LMCJAC}) and its asymptotical extension when ($\alpha\rightarrow \infty; \beta \rightarrow \infty; \mbox{fixed}\, n$) allow us to find the pseudo-classical limit of the physical LMC complexity; they correspond to the high-dimensional or pseudoclassical states. This is because the wavefunctions of such extreme states involve polynomials orthogonal with respect to a Jacobi weight function where both parameters $\alpha$ and $\beta$ are directly proportional to the space dimensionality of the system.

\section{Conclusions and open problems} \label{Conclud}

In this work we have determined the Cram\'er-Rao, Fisher-Shannon and LMC complexity-like measures of the Jacobi polynomials $\hat{P}_n^{(\alpha,\beta)}(x)$, with $\alpha,\beta>-1$, in the extreme situations ($n\rightarrow \infty$; fixed $\alpha,\beta$) and  ($\alpha\rightarrow \infty$; fixed $n,\beta$). They are given by the leading term of the degree and parameter asymptotics of the corresponding statistical properties of the associated probability density (Rakhmanov's density), respectively. Each of these complexity quantifiers capture in a simultaneous way two polynomial's configurational facets of dispersion (variance) and entropic (Fisher, Shannon) types. Briefly, in the limit ($n\rightarrow \infty$; fixed $\alpha,\beta$) we have found that both Cram\'er-Rao and Fisher-Shannon complexities follow a qualitatively similar $n^3$-law behavior for all $\beta$ (see Table  \ref{Table_1}), but the LMC complexity has a different asymptotical $n$-behavior depending on $\beta$ (see Table  \ref{Table_2}). Moreover, in the limit ($\alpha\rightarrow \infty$; fixed $n,\beta$) we have found that  the Fisher information and the variance follow a direct and inverse quadratic dependence on $\alpha$, respectively, while the second-order entropic moment and the Shannon entropy power follow a direct and inverse linear dependence on $\alpha$. The combination of the two dispersion/entropic factors involved for the Cram\'er-Rao, Fisher-Shannon and LMC complexities lead to a constant leading term  for all the complexity measures.\\

Finally, a number of open related problems can be highlighted. First, the extension of these results to the varying Jacobi polynomials \cite{Kuijlaars2004, Kuijlaars2005,Buyarov1999} (i.e., when the parameters depend on the polynomial degree) as well as to the exceptional Jacobi polynomials \cite{Chaitanya2012,Sasaki2010,Quesne2008}, which are very useful to standard and supersymmetric quantum mechanics \cite{Gangopadhyaya2017}. Second,  the determination of the general statistical complexity measures \cite{Sobrino2017} of Fisher-R\'enyi \cite{Nagy2012,Puertas2017c,Toranzo2017,Romera2008,Romera2009} and LMC-R\'enyi \cite{Pipek1997,Lopez2009,Romera2008,Nath2021,Zozor2017} types for the standard and varying Jacobi polynomials; this includes the calculation of the R\'enyi entropy of such polynomials. These open issues are not only interesting \textit{per se} but also because of their chemical and physical applications, especially for the extreme quantum states of highly excited Rydberg and high dimensional types of numerous atomic and molecular systems whose bound states are described by wavefunctions controlled by these polynomials.

\section*{Funding information}
{The work of N. Sobrino has been partially supported by the Basque Government and UPV/EHU under grant IT1249-19. The work of J. S. Dehesa has been partially supported by the Agencia Estatal de Investigaci\'on (Spain) and the European Regional Development Fund (FEDER) under the grant PID2020-113390GB-I00, and by Agencia Andaluza del Conocimiento of the Junta de Andaluc\'ia (Spain) under grant PY20-00082.}
%


\begin{thebibliography}{88}
\bibitem{Sen2011} K.D. Sen Ed., \textit{Statistical Complexities: Applications in Electronic Structure}, Springer, Berlin, \textbf{2011}.
\bibitem{Antolin2009} Antol\'in, J. Angulo, J.C. {\em Int. J. Quant.~Chem.} {\bf 2009}, {\em 109}, 586.
\bibitem{Badii1997}  R. Badii, A. Politi,  \emph{Complexity: Hierarchical Structure and Scaling in Physics}, Henry Holt, New York \textbf{1997}.
\bibitem{Shiner1999} J. S. Shiner, M. Davison,  P. T. Landsberg, {\em Phys. Rev. E} {\bf 1999}, {\em 59}, 1459.

\bibitem{Nielsen2000} M.A. Nielsen, I.L. Chuang, \textit{Quantum Computation and Quantum Information}, 2nd ed., Cambridge University Press, Cambridge \textbf{2000}.
\bibitem{Bruss2019} D. Bruss, G. Leuchs,  \textit{Quantum Information: From Foundations to Quantum Technology} Wiley-VCH, Weinheim, \textbf{2019}.

\bibitem{Cubitt2014} T. Cubitt, \textit{Advanced Quantum Information Theory}, Cambridge University Press, Cambridge, \textbf{2014}.

\bibitem{Vitanyi2001} P.M.B. Vitanyi, IEEE Transactions on Information Theory, Vol. 47, 6, 2001
\bibitem{Lloyd2001}  S. Lloyd, {\em IEEE Control Syst. Mag.} {\bf 2001}, 21, 7.


\bibitem{Dehesa2006} J.S. Dehesa, P. S\'anchez-Moreno, R.J. Y\'a\~nez, J. Comput. Appl. Math. \textbf{2006}, 186, 523.

\bibitem{Dehesa2015} J.S. Dehesa, A. Guerrero and P. S\'anchez-Moreno, J. Comput. Appl. Math, \textbf{2015}, 284, 144.
\bibitem{Ghosh2019} P. Ghosh, D. Nath,  Int. J. Quantum Chem. \textbf{2019}, 119, e25964.

\bibitem{Rudnicki2016} L. Rudnicki, I.V. Toranzo, P. S\'anchez-Moreno and J.S. Dehesa, Phys. Lett. A \textbf{2016}, 380, 377.
 \bibitem{Nikiforov1988} A.F. Nikiforov, V.B. Uvarov, \textit{Special Functions of Mathematical Physics}, Birkh\"auser, Basel, \textbf{1988}.

\bibitem{Goldreich2008} O. Goldreich, \textit{Computational Complexity: A Conceptual Perspective}, Cambridge University Press, Cambridge, \textbf{2008}.
\bibitem{Kaltchenko2021} A. Kaltchenko, arXiv:2109.01960v1 (5 September \textbf{2021})
\bibitem{Parr1989} R. G. Parr, W. Yang, \textit{Density Functional Theory of Atoms and Molecules}. Oxford Univ. Press: Oxford, \textbf{1989}.

\bibitem{Vignat2003}  C. Vignat, J.F. Bercher, {\em Phys. Lett. A} {\bf 2003}, {\em 312}, 27.

\bibitem{Romera2004} E. Romera, J.S. Dehesa, J. Chem. Phys. \textbf{2004}, 120, 8906.

\bibitem{Angulo2008}  J.C. Angulo, J. Antol\'in, K.D. Sen, Phys. Lett. A \textbf{2008}, 372, 670. 
\bibitem{Lopezruiz1995}  R. L\'opez-Ruiz,  H.L. Mancini, X.  Calbet, {\em Phys. Lett. A} {\bf 1995}, 209, 321.
\bibitem{Catalan2002} R.G. Catalan, J. Garay, R. L\'opez-Ruiz,  Phys. Rev. E \textbf{2002}, 66. 011102.

\bibitem{Lopez2005}  R. L\'opez-Ruiz, {\em Biophys. Chem.} {\bf 2005}, {\em 115}, 215.


\bibitem{Pipek1997} J. Pipek, I. Varga, {\em Int. J. Quant. Chem.} {\bf 1997}, {\em 64}, 85.
\bibitem{Romera2009} E. Romera,  R. L\'opez-Ruiz,  J. Sa\~nudo, A. Nagy, {\em Int. Rev. Phys.} {\bf 2009}, {\em 3}, 207.
\bibitem{Lopez2009} R. L\'opez-Ruiz, A. Nagy, E. Romera, J. Sa\~nudo, {\em J. Math. Phys.} {\bf 2009}, 50, 123528. 

\bibitem{Sanchez2014} P. S\'anchez-Moreno, J.C. Angulo, J.S. Dehesa, {\em Eur. Phys. J. D} {\bf 2014}, {\em 68}, 212.
\bibitem{Romera2008}  E. Romera, A. Nagy, {\em Phys. Lett. A} {\bf 2008}, {\em 372}, 6823.
\bibitem{LopezRosa2009} S. L\'opez-Rosa, D. Manzano, J.S. Dehesa,  Physica A \textbf{2009}, 388, 3273.

\bibitem{Martin2003}  M.T. Martin, A. Plastino, O.A. Rosso, {\em Phys. Lett. A} {\bf 2003}, {\em 311}, 126.
\bibitem{Toranzo2017} I.V. Toranzo, P. S\'anchez-Moreno, {\L}. Rudnicki, J. S. Dehesa, Entropy \textbf{2017}, 19, 16
\bibitem{Puertas2017c} D. Puertas-Centeno, I.V. Toranzo, J. S. Dehesa, J. Statist. Mech.: Th. Exp. \textbf{2017}, 1704(4), 043408
\bibitem{Puertas2017d} D. Puertas-Centeno, I.V. Toranzo and J. S. Dehesa, J. Physics A: Theor. Math. \textbf{2017}, 50, 505001
\bibitem{Martin2006} M. Martin, A. Plastino, O. Rosso, J. Math. Chem. \textbf{2006} 369, 439.
\bibitem{Dehesa2009} J.S. Dehesa, S. Lopez-Rosa, D. Manzano, Eur. Phys. J. D \textbf{2009}, 55, 539.
\bibitem{Molina2012} M. Molina-Esp\'iritu, R.O. Esquivel, J.C. Angulo, J. Antol\'in, J.S. Dehesa, J. Math. Chem. \textbf{2012}, 50, 1882.
\bibitem{Angulo2011} J.C. Angulo, J. Antol\'in, R.O. Esquivel, Atomic and molecular complexities: Their physical and chemical interpretations, in: K.D. Sen (Ed.), Statistical Complexities: Applications in Electronic Structure, Springer, Berlin, \textbf{2011}.
\bibitem{Dehesa2011} J.S. Dehesa, S.Lopez-Rosa, D. Manzano, Entropy and complexity analyses of $d$-dimensional quantum systems, in: K.D. Sen(Ed.), Statistical Complexities: Applications in Electronic Structure, Springer, Berlin, \textbf{2011}.
 \bibitem{Bagrov1990} Bagrov, V.G., Gitman, D.M.: Exact Solutions of Relativistic Wavefunctions, Kluwer Acad. Publ., Dordrecht \textbf{1990}.
\bibitem{Cooper2001} F. Cooper, A. Khare, U. Sukhature, \textit{Supersymmetry in Quantum Mechanics}, World Sci. Publ., London \textbf{2001}.
\bibitem{Gangopadhyaya2017} A. Gangopadhyaya, J. Mallow, C. Rasinariu, \textit{Supersymmetric Quantum Mechanics: An Introduction}, World Scientific, Singapore, \textbf{2017}.
\bibitem{Chaitanya2012} K. V. S. S. Chaitanya, S. S. Ranjani, P. K. Panigrahi, R. Radhakrishnan, V. Srinivasan, Pramana \textbf{2015}, 85(1), 53.
\bibitem{Yanez1994} R.J. Y\'a\~nez, W. Van Assche, J.S. Dehesa, \textit{Phys. Rev. A} {\bf 1994}, 50, 3065.

 \bibitem{Andrews1999} G.E. Andrews, R. Askey, R. Roy, \textit{Special functions}, Encyclopedia for Mathematics and Its Applications 16, Cambridge: Cambridge University Press, \textbf{1999}.
  \bibitem{Ismail2005} M.E.H. Ismail,  \textit{Classical and Quantum Orthogonal Polynomials in One Variable}, Encyclopedia for Mathematics and Its Applications, Cambridge University Press, Cambridge \textbf{2005}.
  

 \bibitem{Olver2010} F.W.J. Olver, D.W. Lozier, R.F. Boisvert, C.W. Clark, \textit{NIST Handbook of Mathematical Functions}, Cambridge University Press, New York, \textbf{2010}.
 
 \bibitem{Koekoek2010} R. Koekoek, P. A. Lesky, R. F. Swarttouw, \textit{Hypergeometric Orthogonal Polynomials and their $q$-Analogues}, Springer, Berlin, \textbf{2010}.
 \bibitem{Dehesa2021} J.S. Dehesa, Symmetry \textbf{2021}, 13(8), 1416.

\bibitem{Sobrino2021} J.S. Dehesa, N. Sobrino, Complexity-like properties and parameter asymptotics of $L_q$-norms of Laguerre and Gegenbauer polynomials. J. Phys. A: Math. Theoret. \textbf{2021}

\bibitem{Alhaidari2021} A. D. Alhaidari, I. A. Assi, arXiv:2109.05069v1 (10 September \textbf{2021})

\bibitem{Assi2021} I.A. Assi, A. D. Alhaidari, H. Bahlouli,   J. Math. Phys. \textbf{2021}, 62, 093501

\bibitem{Nath2021} D. Nath, Int. J. Quantum Chem. (2021) e26816
\bibitem{Jacobi1826}C.G.J. Jacobi, J. Reine und Angewandte Mathematik \textbf{1826}, 1, 301.
\bibitem{Jacobi1859} C.G.J. Jacobi, J. Reine und Angewandte Mathematik \textbf{1859}, 56, 149.
\bibitem{Szego1939} G. Szeg\"o, \textit{Orthogonal Polynomials}, American Mathematical Society, Providence, Rhode Island, \textbf{1939}.
\bibitem{Yang2021} X.J. Yang, \textit{An Introduction to Hypergeometric, Supertrigonometric, and Superhyperbolic Functions}, Elsevier, New York, \textbf{2021}. 

 \bibitem{Rakhmanov1977} E. A. Rakhmanov, Math.  USSR-Sb., \textbf{1977}, 32(2), 199.
\bibitem{Buyarov2004} V. Buyarov, J.S. Dehesa, A. Mart\'inez-Finkelshtein, J. S\'anchez-Lara, SIAM J. Sci. Comput. \textbf{2004}, 26, 488.
\bibitem{Perrey2009} E. Perrey-Debain, I. D. Abrahams, SIAM J. Sci. Comput. \textbf{2009}, 31, 3884.
 
\bibitem{Dembo1991} A. Dembo, T.M. Cover, J.A. Thomas, IEEE Trans. Inform. Theory \textbf{1991}, 37, 1501.
\bibitem{Fisher1925} R.A. Fisher, Theory of statistical estimation. \textit{Proc. Cambridge Phil. Soc.} \textbf{1925}, 22, 700,. Reprinted in J.H. Bennet (Ed.) \textit{Collected Papers of R.A. Fisher} University of Adelaide Press: Adelaide, \textbf{1972}, pp. 15-40.
\bibitem{Frieden2004} B.R. Frieden, \textit{Science from Fisher information}, Cambridge University Press, Cambridge,  \textbf{2004}.
\bibitem{SanchezRuiz2005} J. S\'anchez-Ruiz, J.S. Dehesa, J. Comput. Appl. Math. \textbf{2005}, 182, 150.
\bibitem{Guerrero2010} A. Guerrero, P. S\'anchez-Moreno, J.S. Dehesa, J. Phys. A: Math. Theor. \textbf{2010}, 43, 305203.
\bibitem{Sanchez2000} J. S\'anchez-Ruiz, J.S.  Dehesa, J. Comput. Appl. Math. \textbf{2000}, 118, 311.
\bibitem{Aptekarev1994} A.I. Aptekarev, V. Buyarov, J.S. Dehesa,  Russian Acad. of Sci. Sbornik Math. \textbf{1994}, 185(8), 3; English translation Russian Acad. Sci. Sb. Math. \textbf{1995}, 82(2), 373.
\bibitem{Aptekarev1996} A. I. Aptekarev, V. S. Buyarov, W. van Assche, and J. S. Dehesa, Dokl. Math. {\bf 1996},  53, 47.

\bibitem{Dehesa2001} J.S. Dehesa, A. Martinez-Finkelshtein, J. Sanchez-Ruiz, J. Comput. Appl. Math.  {\bf 2001}, 133, 23.

\bibitem{Aptekarev2021} A.I. Aptekarev, E. Belega, J.S. Dehesa, J. Phys. A: Math. Theor. \textbf{2021}, 54, 035305.


\bibitem{Kuijlaars2005} A.B.J. Kuijlaars, A. Mart\'inez-Finkelshtein and R. Orive, Electronic Transactions on Numerical Analysis \textbf{2005}, 19, 1.
\bibitem{Kuijlaars2004} A.B.J. Kuijlaars, A. Mart\'inez-Finkelshtein,  J. d'Analyse Mathematique \textbf{2004}, 94, 195.
\bibitem{Buyarov1999} V.S. Buyarov, J.S. Dehesa, A. Mart\'inez.Finkelshtein, E. B. Saff, J. Approx. Theory, \textbf{1999}, 99, 153.

\bibitem{Sasaki2010} R. Sasaki, S. Tsujimoto, and A. Zhedanov, J. Phys. A: Mathematical and Theoretical \textbf{2010}, 43, 315204.
\bibitem{Quesne2008} C. Quesne, J. Phys. A: Math. Theor. \textbf{2008}, 41, 392001

\bibitem{Sobrino2017} N. Sobrino-Coll, D Puertas-Centeno, I. V. Toranzo, J. S. Dehesa, J. Statist. Mechanics: Theory and Exp. \textbf{2017}, 083102.
\bibitem{Zozor2017} S. Zozor, D. Puertas-Centeno, J.S. Dehesa, Entropy \textbf{2017}, 19, 493

\bibitem{Nagy2012} A. Nagy, E. Romera, Physica A \textbf{2012}, 391, 3650


%
%
%
%
%
%
%
%
%
%
%
%
%
%
%
%
%
%
%
%
%
%
%
%
%
%
%
%
%
%
%
%
%
%
%
%
%
%
%

 


\end{thebibliography}
\end{document}